\documentclass{scrartcl}
\usepackage[T1]{fontenc}
\usepackage[latin1]{inputenc}
\usepackage{graphics}
\usepackage{subfigure}

\title{Diffractive Dijet Production}
\usepackage{amsmath}
\usepackage{verbatim}
\usepackage{graphicx}
\usepackage{amsfonts}

\author{R.B. Appleby and J.R. Forshaw \\ Department of Theoretical Physics \\ The University of Manchester \\  Manchester \\
 M13 9PL \\ United Kingdom }
\date{}

\begin{document}

{\par\raggedleft \texttt{MC-TH-01/08} \par}
\vspace{1.5cm}

{\par\centering {\LARGE Diffractive Dijet Production} {\huge }\huge \par}
\bigskip{}

{\par\centering {\large R.B. Appleby and J.R. Forshaw }\large \par}

{\par\centering Department of Physics \& Astronomy \\
 University of Manchester \\
 Manchester M13 9PL \\
 United Kingdom \par}

\begin{abstract}
We explore the diffractive interaction of a proton with an anti-proton which
results in centrally produced dijets. This process has been recently studied
at the Tevatron. We make predictions within an Ingelman-Schlein approach 
and compare them to the recent data presented by the
CDF collaboration. Earlier calculations resulted in theoretical 
cross-sections which are much larger than those observed by CDF. We find 
that, after consideration of hadronisation effects and the parton shower, and
using parton density functions extracted from diffractive deep inelastic
scattering at HERA, it is possible to explain the CDF data. We need to assume 
a gap survival probability of around 10\% and this is in good agreement with
the value predicted by theory. We also find that the non-diffractive 
contribution to the process is probably significant in the kinematical
region probed by the Tevatron.
\end{abstract}

\section{{\large Introduction}\large }

In this work we study the process \( p\bar{p}\rightarrow p+JJX+\bar{p} \),
where \( JJX \) denotes a centrally produced cluster of hadrons containing
at least two jets. The CDF collaboration at FNAL has presented data on this
process \cite{cdf2000}. Such an interaction is diffractive since the colliding
hadrons emerge intact with only a small loss of longitudinal momentum. Events
are characterised by the production of two jets in the central region, 
separated from the final state hadrons by rapidity gaps. Therefore they are 
also known as ``gap-jet-gap'' events. 

Previous calculations of this process \cite{berera2000,acw}
have been carried out at the parton level and have overestimated the total
dijet production cross-section by several orders of magnitude.
In this paper we see that by including the effects of hadronisation
and by using the diffractive parton density functions extracted by the H1
collaboration \cite{adloff1997}, it is possible to fit the  
data in a natural
way. In particular, we shall show that the large suppression which arises from the hadronisation
corrections means that we are able to fit the data with a ``gap survival 
probability'' of around 10\% that is consistent with theoretical expectations 
\cite{kaidalov2001,gotsman1999}.

There are two approaches to central dijet production that have been 
considered in the literature:

\begin{enumerate}
\item The factorised model. In this approach one assumes that factorisation 
\cite{collins98} can be applied to the process, just as in inclusive hard 
scattering processes, modulo an overall multiplicative factor which has
come to be called the ``gap survival probability'' \cite{bjorken1993}.
Following Ingelman and Schlein (IS), one can go further and assume so-called
{}``Regge Factorisation{}'' \cite{ingelmann1985} which assumes that the process
is driven by pomeron exchange, as illustrated in Figure 1. Thus one assigns
an experimentally determined parton density and flux factor to the exchanged
pomeron \cite{adloff1997}. It is expected that this factorisation will be 
violated at the Tevatron \cite{berera2000,collins98,collins1993,martin1997} and the
extent to which the breaking can be accommodated, at some level, by a gap 
survival factor is very much an open question. Central 
dijet production within a factorised model has been studied before 
\cite{acw,berera2000,boonekamp2001}.
\item The non-factorised model. These models have been developed in 
\cite{collins1993,berera1996,khoze2001,kms}.
They are exclusive in
the sense that all energy lost by the interacting hadrons goes into the final
state dijets. Therefore we will see no pomeron remnants in the final state,
as illustrated in Figure 2\footnote{In \cite{kms} the possibility that there is
additional radiation into the final state from the gluons in fig. 2 is
considered.}. This feature would manifest itself as a peak in
the distribution showing the fraction of the available centre-of-mass energy
which goes into the jets; a feature which is absent in the CDF data. However
this need not be the case for Tevatron Run II which ought to be a good place
to look for evidence of a non-factorised contribution. Subsequently, we 
shall not consider such models.
\end{enumerate}

\begin{figure}
{\par\centering \resizebox*{!}{0.3\textheight}{\includegraphics{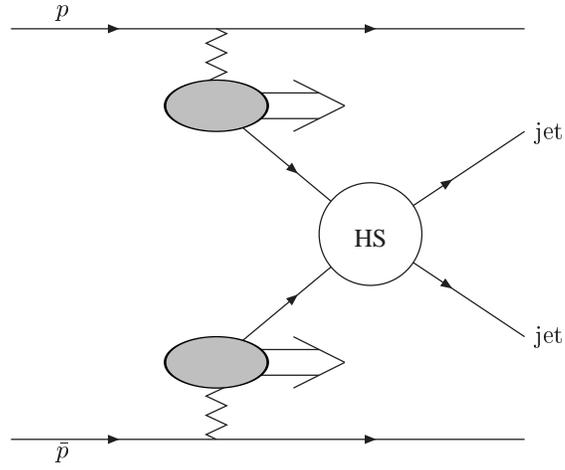}} \par}

\caption{\textit{Dijet production in the factorised model. The zig-zag lines denote
the exchanged pomerons.}}
\end{figure}

\begin{figure}
{\par\centering \resizebox*{!}{0.3\textheight}{\includegraphics{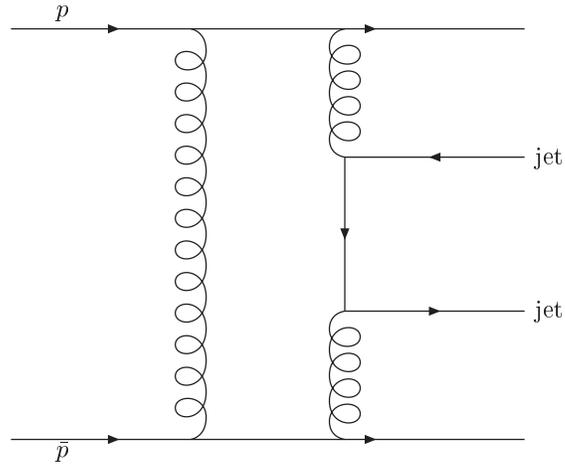}} \par}

\caption{\textit{Dijet production in the non-factorised model.}}
\end{figure}

We use the IS approach in this paper and compare the predictions of the Monte
Carlo event generator, POMWIG \cite{pomwig}, with the experimental results \cite{cdf2000}. 

The experiments were performed at \( \sqrt{s}=1.8 \) TeV and diffractive events
are selected by applying cuts on the hadronic momentum loss. The details of
the experiment are described in \cite{cdf2000} and the relevant cuts are:

\begin{itemize}
\item The anti-proton fractional energy loss \( \xi _{\bar{p}} \) satisfies \( 0.035<\xi _{\bar{p}}<0.095 \).
Such a condition corresponds to a rapidity gap on the anti-proton side. This
cut is made by tagging the anti-proton. 
\item The proton fractional energy loss \( \xi _{p} \) satisfies \( 0.01<\xi _{\bar{p}}<0.03 \).
Such a condition corresponds to a rapidity gap on the proton side. Note that
the proton is not tagged at CDF and the cut is made by looking for a rapidity
gap on the proton side. 
\item \( |t_{\bar{p}}|<1 \) GeV\( ^{2} \). \( |t_{\bar{p}}| \) is the four-momentum
transfer from the anti-proton squared. 
\item The event must have two or more jets in the pseudo-rapidity region \( -4.2<\eta <2.4 \)
and at least two jets must have a minimum transverse energy of \( E_{T}^{min}=7 \)
GeV or 10 GeV.
\end{itemize}
The final state jets were found using the cone algorithm with a cone radius
of 0.7 and an overlap parameter of 0.5.

\section{{\large The Model}\large }

With reference to Figure 1, we assign the pomeron a partonic content and the
dijet production is driven by the perturbative QCD interaction of partons from
the colliding pomerons. The production cross-section for this so-called {}``Double
Pomeron Exchange{}'' process (DPE) can be written \\
\begin{align}
\frac{d\sigma_{DPE}^{dijet}}{d\eta_3 d\eta_4 dp_{\perp}^2}(p\bar{p} \to 
p+JJX+\bar{p}) = \int d\xi_p \int d\xi_{\bar{p}} F_{\mathbb{P}/p}(\xi_p) F_{\mathbb{P}/\bar{p}}(\xi_{\bar{p}}) \nonumber \\
\sum \beta_p f_{i/\mathbb{P}}(\beta_p) \beta_{\bar{p}} f_{j/\mathbb{P}}(\beta_{\bar{p}}) \frac{d\hat{\sigma}_{HS}}{d\hat{t}}(ij \to kl)
\end{align} \\
where \( F_{\mathbb {P}/p}(\xi )=F_{\mathbb {P}/\overline{{p}}}(\xi ) \) is
the pomeron flux factor, \( \beta  \) is the fraction of the pomeron momentum
carried by the parton entering the hard scattering and \( f_{i/\mathbb {P}}(\beta ) \)
is the pomeron parton density function for partons of type \( i \). The rapidity
of the outgoing partons is denoted \( \eta _{3} \) and \( \eta _{4} \),
their transverse momentum is \( p_{\perp } \) and $\frac{d\hat{\sigma}_{HS}}{d\hat{t}}(ij \to kl)$
denotes the QCD 2-to-2 scattering amplitudes. 
The parton transverse momentum, $p_{\perp }$ is equal to
the jet transverse energy $E_T$ at the parton level.

Note that the dijets are accompanied by pomeron remnants which we observe in
the final state alongside the jets and the diffracted hadrons. We use the H1
leading order pomeron fits to \( F_{2}^{D(3)} \) measurements 
\cite{adloff1997}.
The measurement of \( F_{2}^{D(3)} \) is quark dominated and gluon sensitivity
only enters through scaling violations. Hence the gluon density has quite a
large uncertainty. This is important for the gluon dominated DPE process. 
The gluon densities of the H1 fits are illustrated in Figure 3. The mean 
value of  $\beta$ relevant at the Tevatron is in the region of 0.3 to 0.4.
Fit 4 has a gluon content that is heavily suppressed relative to fits 5 and
6, and fit 6 is peaked at high \( \beta  \). Fits 5 and 6 are now the favoured
fits to describe H1 data.

\begin{figure}
{\par\centering \subfigure{\resizebox*{0.6\columnwidth}{!}{\includegraphics{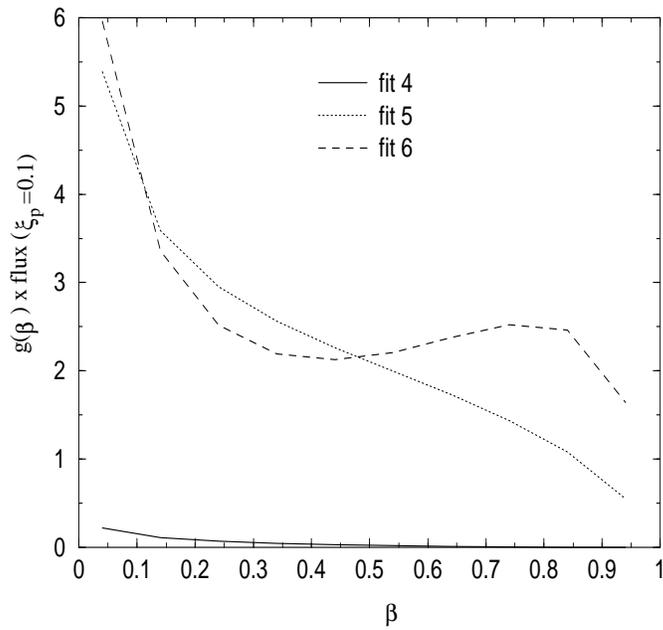}}} \par}

\caption{\textit{The gluon densities in the Pomeron as fitted by H1. The fits are evaluated
at \protect\( Q^{2}=50\: GeV^{2}\protect \), the typical scale of the jet transverse
energy at CDF, and are multiplied by the Pomeron flux evaluated at \protect\( \xi _{p}=0.1\protect \). }}
\end{figure}

In our analysis we also include the effect of non-diffractive contributions
through an additional regge exchange, which we refer to as the reggeon 
contribution. This is expected to be important
in the region of $\xi_{\bar{p}}$ explored at the Tevatron. Following
H1, we estimate reggeon exchange by assuming that the reggeon can be 
described by the pion parton densities of Owens \cite{Owens}. 
This contribution is added incoherently to the pomeron contribution.
For more details of the implementation we refer to \cite{pomwig}.

\section{{\large Results}\large }

\subsection{Hadronisation \& Parton Shower Effects}
Figure 4 demonstrates the agreement between POMWIG and an independent 
calculation
for the total DPE cross-section as a function of the minimum jet transverse
energy \( E^{min}_{T} \). These curves were produced using H1 fit 5 and contain
only the pomeron exchange contribution.
More interestingly, we see that the curve which describes the total 
cross-section after the effects of hadronisation and parton showering 
have been included shows
a significant reduction relative to the naive parton level calculation. This
suppression effect may be understood by observing that the effect is a shift
in the cross-section by \( \Delta E_{T}=2 \) GeV. This is a direct consequence
of the broadening of the jet profile by the parton shower and hadronisation.
The reduction is lower for the quark dominated fit (H1 fit 4) since quark jets
tend to have a narrower profile. 
\begin{figure}
{\par\centering \resizebox*{0.75\columnwidth}{!}{\includegraphics{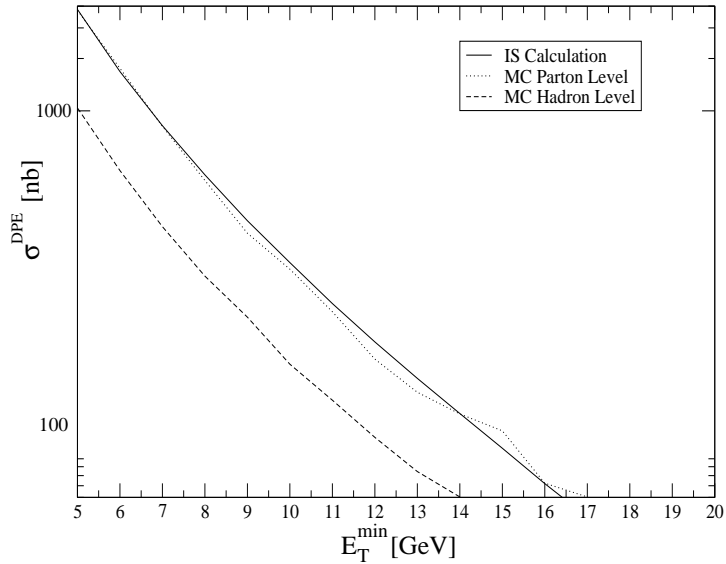}} \par}
\caption{\textit{The total cross-section at both parton and hadron level, plotted as
a function of the minimum transverse energy of the dijets. }}
\end{figure}
\subsection{Total Cross-Section}
Now we turn our attention to the comparison of our theoretical predictions to
the measured total cross-section. The analysis we perform includes the processes where the
exchanges are either both pomerons or both reggeons. We do not include the case where one
is a reggeon and the other is a pomeron, nor do we include interference contributions.
Whilst the latter may be small, the former will not be if the pure reggeon
contribution is not negligible. This limitation arises since pomeron-reggeon interactions 
are not yet included in POMWIG. The results
are presented in Tables 1 and 2 (for the data the first error is statistical
and the second is systematic).

{\centering \begin{tabular}{|c|c|c|}
\hline 
&
 Parton Level {[}nb{]} &
 Hadron Level {[}nb{]} \\
\hline 
CDF Result &
&
 43.6 \( \pm  \) 4.4 \( \pm  \) 21.6\\
\hline 
I\( \!  \)P fit 4 &
 6 &
 2.7 \\
\hline 
I\( \!  \)P fit 5 &
 815 &
 230 \\
\hline 
I\( \!  \)P fit 6 &
 1175 &
 339 \\
\hline 
I\( \!  \)R &
 241 &
 58 \\
\hline 
I\( \!  \)P+I\( \!  \)R fit 4 &
 247 &
 60.7 \\
\hline 
I\( \!  \)P+I\( \!  \)R fit 5 &
 1056 &
 288 \\
\hline 
I\( \!  \)P+I\( \!  \)R fit 6 &
 1416 &
 397  \\
\hline 
\end{tabular}\par}

{\par\centering \textit{Table 1. The total DPE cross-sections for a jet cut of
7 GeV. The contributions from pomeron and reggeon exchange are shown separately. }\par}

{\centering \begin{tabular}{|c|c|c|}
\hline 
&
 Parton Level {[}nb{]} &
 Hadron Level {[}nb{]} \\
\hline 
CDF result &
&
 3.4 \( \pm  \) 1.0 \( \pm  \) 2.0\\
\hline 
fit 4 &
 1 &
 0.47 \\
\hline 
fit 5 &
 123 &
 33 \\
\hline 
fit 6 &
 187 &
 66 \\
\hline 
I\( \!  \)R &
 24 &
 1.7 \\
\hline 
I\( \!  \)P+I\( \!  \)R fit 4 &
 25 &
 2.17 \\
\hline 
I\( \!  \)P+I\( \!  \)R fit 5 &
 147 &
 34.7 \\
\hline 
I\( \!  \)P+I\( \!  \)R fit 6 &
 211 &
 67.7  \\
\hline 
\end{tabular}\par}

{\par\centering \textit{Table 2. The total DPE cross-sections for a jet cut of
10 GeV. The contributions from pomeron and reggeon exchange are shown separately. }\par}

Using fit 5, the overall cross-section that we predict for a jet cut of 7 GeV
is 288 nb which is in excess of the experimental value of 43.6 nb. A similar
excess is present with a cut of 10 GeV. However, we can match our results to
the data if we assume an overall multiplicative gap survival probability of
around 15\%. The large reggeon contribution implies a non-negligible
pomeron-reggeon contribution and naively estimating this as twice the geometric mean
of the pomeron-pomeron and reggeon-reggeon contributions would push the gap
survival factor down to around 10\%. Given that the systematic error on the CDF 
cross-sections is 
high, that the uncertainty in our knowledge of the gluon density
directly affects the normalisation of the cross-section and that the size of
the non-diffractive reggeon contribution is also uncertain it is not possible
to make a more precise statement about gap survival. In any case the value we
obtain agrees well with the expectations of \cite{kaidalov2001,gotsman1999}. 
It should be appreciated that this agreement of our estimate of the 
gap survival probability with other theoretical estimates is only possible 
after the inclusion of hadronisation effects.
Both fits 5 and 6 can describe the data, although measurements of
diffractive dijet production at HERA suggest that fit 5 is favoured \cite{dijets}. Our total cross-section agrees with that of \cite{cox2001}. 
The ratio of the fit 5 to the fit 4 cross-sections is of the order of 100, 
which we can understand from the different gluon densities illustrated in 
Figure 3. Note that the relative size of the non-diffractive contribution 
compared to the diffractive contribution is not small.

We can study further the suppression of the total cross-section, relative to
the naive parton level result, by looking at the particular role of the parton
shower phase in POMWIG. We have performed the total cross-section calculation
for H1 fit 5 after parton showering but before hadronisation. The results are
presented in Table 3, for an \( E^{min}_{T} \) of 7 GeV.

{\centering \begin{tabular}{|c|c|c|c|}
\hline 
&
 Parton Level {[}nb{]} &
 Parton Shower {[}nb{]} &
 Hadron Level {[}nb{]} \\
\hline 
CDF result &
&
&
 43.6 \( \pm  \)4.4 \( \pm  \)21.6\\
\hline 
I\( \!  \)P &
 815 &
 421 &
 230 \\
\hline 
I\( \!  \)R &
 241 &
 144 &
 58 \\
\hline 
I\( \!  \)P+I\( \!  \)R &
 1056 &
 565 &
 288  \\
\hline 
\end{tabular}\par}

{\par\centering \textit{Table 3. The total DPE cross-section at the parton shower
and hadron levels. }\par}

Not surprisingly, the parton shower phase of POMWIG is responsible for a large
part of the suppression relative to the naive parton level prediction.

\subsection{Event distributions}

In Figures 5 and 6 we show distributions in number of events of the mean jet transverse energy,
\( E_{T}^{*} \), the mean jet rapidity, \( \eta ^{*} \), the azimuthal separation
of the jets, \( \Delta \phi  \), and the dijet mass fraction, \( R_{jj} \):
\begin{align}
R_{jj}=\frac{\sum_i E_i}{\xi_{p}\xi_{\bar{p}}s} \nonumber \\
\approx \beta_p\beta_{\bar{p}}.
\end{align}
The sum in the numerator is over all particles in the dijets.
Some of these distributions have also been examined in \cite{boonekamp2001}.
In Figure 5 we compare the data to results at the parton and hadron level, and
we show the reggeon contribution separately\footnote{All curves except the 
reggeon are area normalised to unity. The reggeon is normalised
relative to the total.}. In Figure 6 we show results at the hadron level 
for the three different H1 pomeron parton density functions.

\begin{figure}
{\par\centering \resizebox*{0.95\columnwidth}{!}{\includegraphics{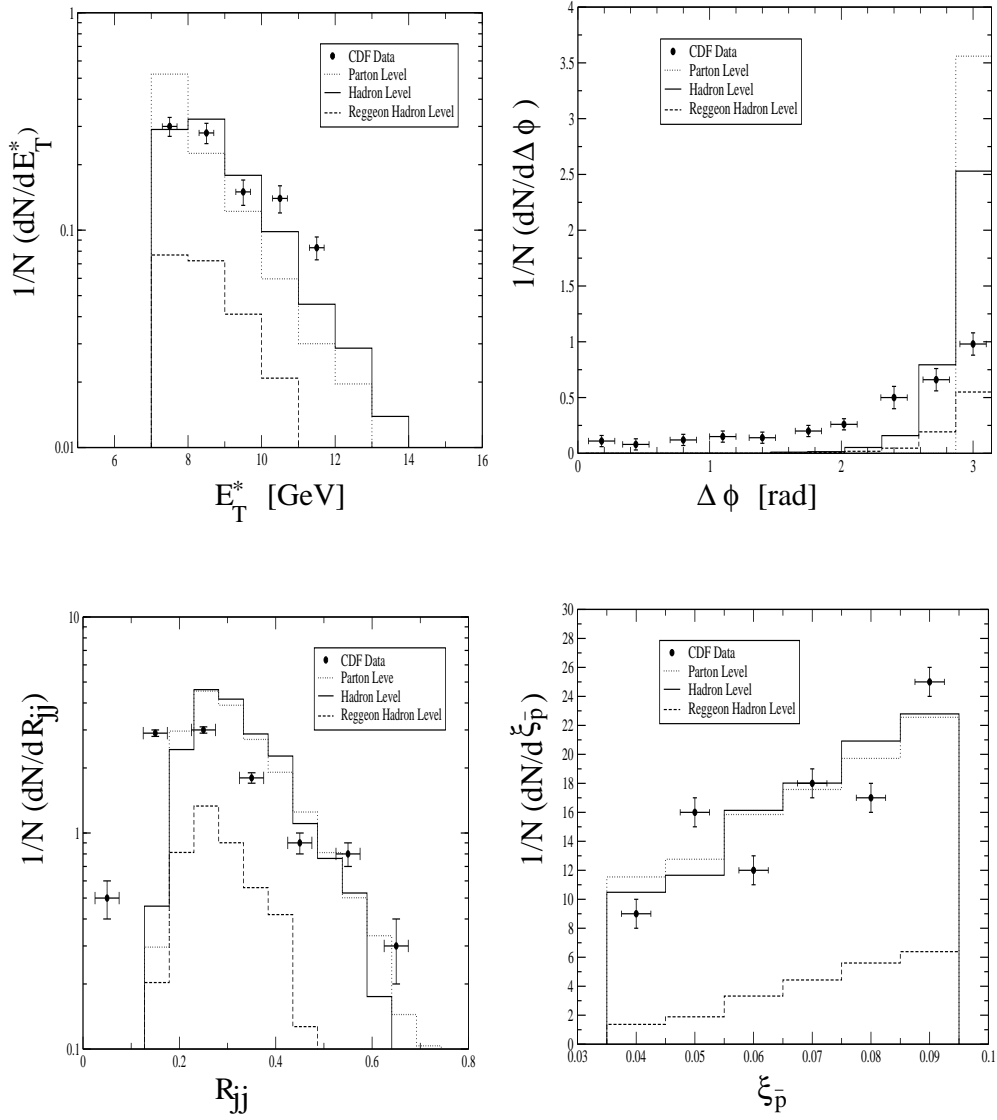}} \par}

\caption{\textit{Comparison of theoretical predictions at the parton and the hadron
level. Also shown is the separate contribution from the reggeon (normalised
relative to the total).}}
\end{figure}

\begin{figure}
{\par\centering \resizebox*{0.95\columnwidth}{!}{\includegraphics{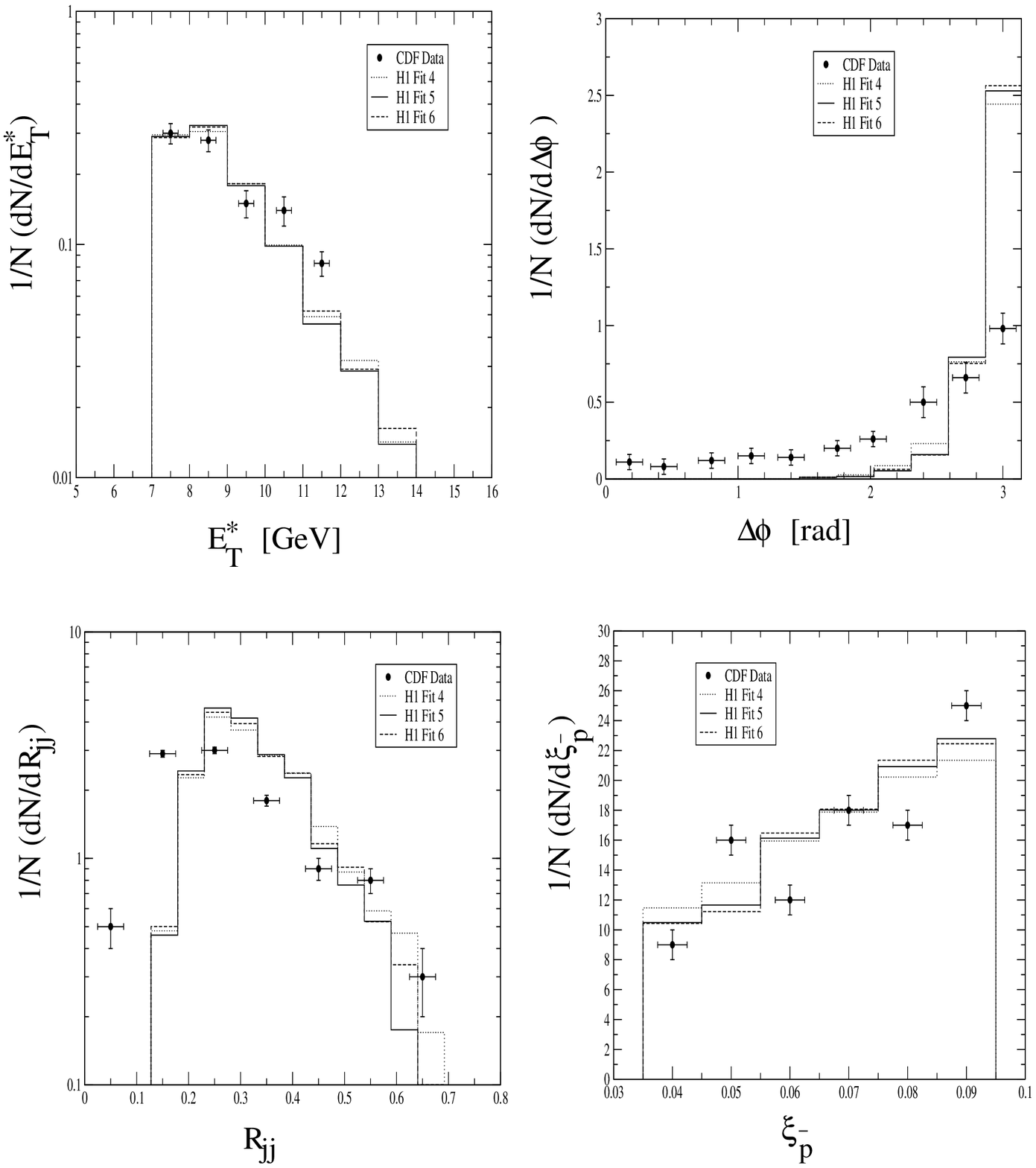}} \par}

\caption{\textit{Comparison of theoretical predictions for the different H1 fits to
the pomeron parton density functions at the hadron level.}}
\end{figure}

We urge caution when comparing the data and theory as is done in Figures 5 and
6 since the data are not corrected for detector 
effects\footnote{Primarily because of the low \( E_{T} \) of the CDF jets.
}. The existence of the long tail to low angles in the \( \Delta \phi  \) 
distribution
illustrates the dangers: there is no possibility to produce such a long tail
in a hadron level Monte Carlo simulation. We have explored the effect of smearing
the momenta of all outgoing hadrons and the qualitative effect is to flatten
the \( E^{*}_{T} \) distribution, enhance the \( \Delta \phi  \) tail and
soften the \( R_{jj} \) distribution. Nevertheless, we are unable to draw any
strong conclusions until the corrected data become available.

\section{{\large Conclusion}\large }

The work of this paper has focused on the so-called Double Pomeron Exchange
process recently measured at the Tevatron. We have extended previous 
calculations by including the effects of parton showering and hadronisation 
and found that they lead to a suppression of the cross-section relative to 
the naive parton level by a factor of around 4. We also found that, in the 
kinematic region probed by the Tevatron, the effect of non-diffractive 
(reggeon) exchange is probably important. At the present time the issue of 
gap survival is not very well understood. However, it is encouraging that
we are able to describe the data with a gap survival probability of 
around 10\% which is consistent with previous theoretical estimates.

We would like to thank Arjun Berera, Brian Cox, Dino Goulianos, Christophe Royon
and Mike Seymour. This work was supported by the UK Particle Physics and Astronomy
Research Council.

\end{document}